\documentclass[conference]{IEEEtran}
\IEEEoverridecommandlockouts

\usepackage{pgfplots}
\usepackage{setspace} 
\pgfplotsset{compat=1.17}
\usepackage{subcaption}
\usepackage{makecell}

\usepackage{cite}
\usepackage{amsmath,amssymb,amsfonts}
\usepackage{algorithmic}
\usepackage{graphicx}
\usepackage{textcomp}
\usepackage{xcolor}
\def\BibTeX{{\rm B\kern-.05em{\sc i\kern-.025em b}\kern-.08em
    T\kern-.1667em\lower.7ex\hbox{E}\kern-.125emX}}
\begin{document}

\title{ImprovNet - Generating Controllable Musical Improvisations with Iterative Corruption Refinement\\
\thanks{This work was supported by UKRI and EPSRC (grant EP/S022694/1) and by SUTD's Kickstart Initiative (grant SKI 2021\_04\_06). The training carried out in this research utilized AI Industrial Convergence Cluster supported by the Ministry of Science and ICT of Korea, and Gwangju Metropolitan City.}
}

\author{
\IEEEauthorblockN{
Keshav Bhandari\textsuperscript{1}, 
Sungkyun Chang\textsuperscript{1}, 
Tongyu Lu\textsuperscript{2}, 
Fareza R. Enus\textsuperscript{1}, \\
Louis B. Bradshaw\textsuperscript{1}, 
Dorien Herremans\textsuperscript{2}, 
Simon Colton\textsuperscript{1}
}
\IEEEauthorblockA{
\textsuperscript{1}Electronic Engineering and Computer Science Department, Queen Mary University of London, UK \\
Emails: \{k.bhandari, sungkyun.chang, f.r.enus, l.b.bradshaw, s.colton\}@qmul.ac.uk
}
\IEEEauthorblockA{
\textsuperscript{2}Information Systems, Technology, and Design, Singapore University of Technology and Design, Singapore \\
Emails: \{tongyu\_lu, dorien\_herremans\}@sutd.edu.sg
}
}

\maketitle

\begin{abstract}
Despite deep learning's remarkable advances in style transfer across various domains, generating controllable performance-level musical style transfer for complete symbolically represented musical works remains a challenging area of research. Much of this is owed to limited datasets, especially for genres such as jazz, and the lack of unified models that can handle multiple music generation tasks. This paper presents ImprovNet, a transformer-based architecture that generates expressive and controllable musical improvisations through a self-supervised corruption-refinement training strategy. The improvisational style transfer is aimed at making meaningful modifications to one or more musical elements - melody, harmony or rhythm of the original composition with respect to the target genre. ImprovNet unifies multiple capabilities within a single model: it can perform cross-genre and intra-genre improvisations, harmonize melodies with genre-specific styles, and execute short prompt continuation and infilling tasks. The model's iterative generation framework allows users to control the degree of style transfer and structural similarity to the original composition. Objective and subjective evaluations demonstrate ImprovNet's effectiveness in generating musically coherent improvisations while maintaining structural relationships with the original pieces. The model outperforms Anticipatory Music Transformer in short continuation and infilling tasks and successfully achieves recognizable genre conversion, with 79\% of participants correctly identifying jazz-style improvisations of classical pieces. Our code and demo page can be found at https://github.com/keshavbhandari/improvnet.
\end{abstract}


\section{Introduction}
Music style transfer has recently emerged as a fascinating area of generative AI research, offering new possibilities for creating personalized compositions. Altering the melodic, rhythmic, and harmonic aspects of compositions can enable the crafting of music that feels both familiar and creative, tailored to reflect individual preferences or to evoke specific artist and genre styles.

The concept of neural style transfer was pioneered by \cite{gatys2016image}, with the goal of blending ``content'' from one source with the ``style'' characteristics of another. In this framework, content refers to the underlying structure that remains intact, while style encompasses the features transferred from a different input. Since then, neural style manipulation has sparked widespread interest in various modalities. However, the core idea behind style transfer remains consistent across these applications: preserving the content or structure of one input while adopting the stylistic characteristics of another.

In recent years, the concept of ``style transfer'' has evolved, particularly in the realm of music processing. As seen in \cite{nakamura2019unsupervised,cifka2019supervised,pezzat2020many,dwarkani2020unpaired,mukherjee2022composeinstyle}, the definition now encompasses a wider and more general range of style transformations rather than deriving it from one or a few examples. In this context, the goal is to transform an input to match the stylistic characteristics of large datasets. According to \cite{cifka2020groove2groove}, this can also be described as \textit{style conversion} or \textit{translation}. In our work, we adhere to the style conversion definition from the recent literature. 

In this paper, we focus on the generation of expressive performance improvisations in the context of symbolically represented classical and jazz solo piano pieces. An improvisation could be viewed as a special case of style conversion where the generated output either adopts a different style or retains the original one through meaningful modifications to the given content. The goal of improvisation is to introduce changes that maintain a recognizable link to the original, preserving its identity while exploring new variations within its stylistic representations. Hence, the musical content is generated with the aim of reflecting the essence of the original, either within the same style or a distinct one, depending on the user.

We interpret ``style" throughout this paper as genre-specific elements, such as rhythmic, harmonic, and melodic properties that reflect a particular musical genre. On the other hand, ``content'' encompasses the foundational elements specific to a musical piece, including its core melody, rhythm, chord progressions, and structure. According to this, our goal for cross-genre music improvisation is to adapt the content of a music piece to reflect the stylistic musical attributes representative of the target genre while preserving the core musical ideas of the original. For example, a classical to jazz cross-genre improvisation may preserve motivic fragments, tempo, and global form of the original piece while introducing elements such as dissonance, chromatic scales, irregular or syncopated rhythms, jazz harmonies, and melodies. Similarly, our goal for intra-genre improvisation (classical-to-classical or jazz-to-jazz improvisation) is to adhere to the genre of the original piece while modifying its content, generating a new variation that still references the essence of the initial composition. 

Our study addresses a gap in symbolic music research in generating controllable expressive performance style transfer for complete musical works \cite{dai2018music}. This gap arises from challenges such as limited symbolic data, especially in jazz \cite{edwards2023pijama}, the scarcity of expressive polyphonic solo piano jazz datasets, and the lack of well-annotated datasets \cite{bhandari2024motifs}. These issues hinder the training of generative models capable of capturing stylistic variations and performance dynamics. Furthermore, ambiguities in the definition of target genres \cite{crocker1986history} reduce user control over transformations, leading to low-fidelity outputs. Finally, no single symbolic music model integrates tasks like variation creation, harmonization, infilling, and continuation within a unified framework, an essential step to advance style transfer and enabling personalized music generation.

To address these challenges, we propose ImprovNet, a transformer-based architecture \cite{vaswani2017attention} that is pre-trained and fine-tuned with a self-supervised \textit{corruption refinement infilling strategy}. ImprovNet is designed to generate expressive and controllable improvisations for complete musical pieces. The improvisational style comes from modifications to one or more musical elements - melody, harmony, or rhythm of the original composition to emulate the spontaneity and creativity of human improvisation. The extent to which the stylistic elements get modified dictates how recognizable the original content is. In our approach, we allow the user control over the transformation functions and the degree of transformation applied to the musical elements, letting them choose whether the generation remains highly recognizable or takes on a more creative reinterpretation. Due to the nature of the corruption refinement infilling task, as we explain below in the methods section, ImprovNet is also capable of harmonizing a melodic line, performing short prompt continuation and short infilling tasks in addition to generating style-aware improvisations.

\section{Related Work}

Music style transformations come in various forms, depending on how style is defined, how music is represented, and what the model is conditioned on. In this paper, we focus on expressive and controllable improvisational style in the symbolic domain, achieved through modifications to the rhythm, harmony, and melody of a solo piano performance conditioned on the target genre. Previous studies have focused on these tasks separately. For example, \cite{wu2023musemorphose} addresses style transfer for long pop piano pieces, allowing users to customize musical attributes such as rhythmic intensity and polyphony at the bar level. Models such as \cite{hadjeres2017deepbach,huang2019counterpoint} harmonize a melody in the style of Bach chorales. Similarly, \cite{jiang2020counterpoint} proposes FolkDuet to fuse Chinese melodies with musical counterpoint in real time. \cite{tan2020music} introduces FaderNets for generating musical variations by adjusting low-level attributes using latent regularization and feature disentanglement while enabling high-level feature control (e.g., arousal) through a VAE framework. \cite{zhao2021accomontage,wu2024generating} focus on arranging accompaniments for solo piano melodies, offering control over aspects such as voice density. Studies such as \cite{lv2023getmusic,ren2020popmag,zhaostructured,wu2023c2,min2023polyffusion,cifka2020groove2groove} extend the controllable accompaniment arrangement task to multi-track generation. However, all of these approaches use highly quantized piano encodings, limiting the model's ability to produce expressive renditions.

An expressive performance rendition in symbolic music should be capable of human-like characteristics in timing, dynamics, and articulations, which go beyond the rigid quantization of notes. \cite{performance-rnn-2017} was among the first works to generate expressive performance renditions by modelling a stream of midi note events. Recently, generating expressive performance renditions from symbolic score has been the focus of several works such as \cite{jeong2019graph,jeong2019virtuosonet,xiao2024music,borovik2023scoreperformer,maezawa2019rendering,lenz2024pertok,colton2024automatic}. Among these, \cite{lenz2024pertok} is capable of generating controllable and expressive polyphonic piano variations, but is limited to short fragments of music. 

Several works have explored genre style conversion, which is more closely related to this paper. \cite{nakamura2019unsupervised} explores unsupervised monophonic melody style conversion using a statistical framework that integrates a music language model with an edit model, discovering musical styles through unsupervised grammar induction. \cite{brunner2018symbolic} pioneered the use of GANs for genre conversion in symbolic polyphonic piano pieces by adapting the CycleGAN \cite{zhu2017unpaired} framework with genre classifier discriminators. \cite{fu2020transfer} enhanced this approach by incorporating a beta-VAE \cite{higgins2017beta} for better disentanglement of style and content. Recently, \cite{sulaiman2022genre} and \cite{ding2022steelygan} have advanced multi-instrument genre conversion through architectural and loss function-related modifications to the CycleGAN and VAE frameworks. However, these GAN-based approaches face training instability, potential mode collapse, and challenging parameter tuning. Their piano-roll representations also limit expressive performance capabilities and lack user-controlled style conversion intensity, both key aspects of our work. Recent works on infilling and prompt continuation tasks, including \cite{min2023polyffusion,lv2023getmusic,thickstun2023anticipatory,chang2021variable,hadjeres2021piano}, also relate to our model's capabilities.

\section{Methods}

\subsection{Tokenization}

\begin{figure*}[t]
    \centering
    \includegraphics[width=\textwidth]{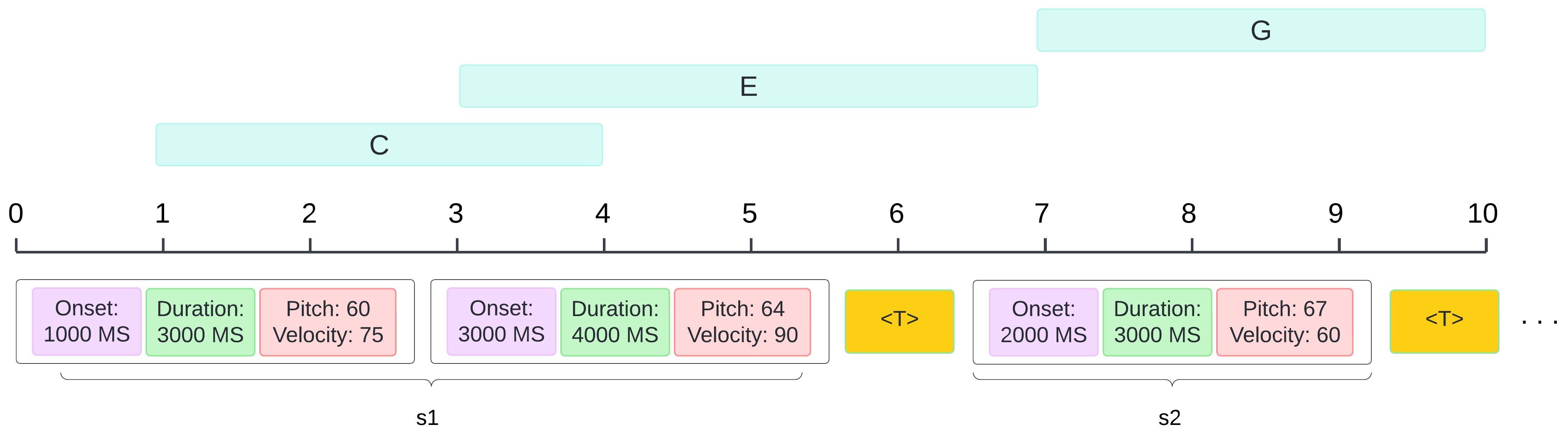}
    \caption{A 10 second piano roll depicting the Aria tokenizer that uses absolute onsets relative to the start of the segment.}
    \label{fig:aria-tokenizer}
\end{figure*}

We adapt a publicly available tokenizer called Aria\footnote{https://github.com/EleutherAI/aria-utils} to encode expressive piano performances for ImprovNet. Although the Aria tokenizer can encode all MIDI instrument categories, we restrict it to solo piano and additionally change the order of tokens to assist with the harmonization task for which the generated logits are constrained, as shown later in section \ref{Harmonization}. As shown in Fig.~\ref{fig:aria-tokenizer}, the Aria tokenizer uses a chunked absolute onset encoding. In this encoding, the pitch and velocity of the MIDI are merged into a single token, while the onset and duration remain as separate tokens. Most metadata, such as bars, key, and time signatures, is excluded. However, essential musical elements such as the usage of the sustain pedal are directly incorporated into the onset and duration tokens. Quantization is minimal: onset and duration values are rounded to the nearest 10 milliseconds, and velocity values are quantized to increments of 15 MIDI units (ranging from 0 to 120). The use of minimally quantized absolute onsets allows the music to be encoded in a manner that closely preserves the characteristics of a human-like performance rendition.

One disadvantage of using absolute onsets is that the vocabulary size scales with the length of the piece. This is problematic as transformer models struggle with arithmetic~\cite{zhou2023algorithms}. To address this issue and avoid sparse tokens during training, the Aria tokenizer chunks the sequence into 5,000 millisecond segments, separating each chunk with a special token $\langle T \rangle$. The $\langle T \rangle$ token resets the absolute onsets to start from 0, similar to the function of the `bar' token in the REMI~\cite{huang2020pop} encoding, thus controlling the vocabulary size.

\begin{figure}[h!]
    \centering
    \includegraphics[width=8cm]{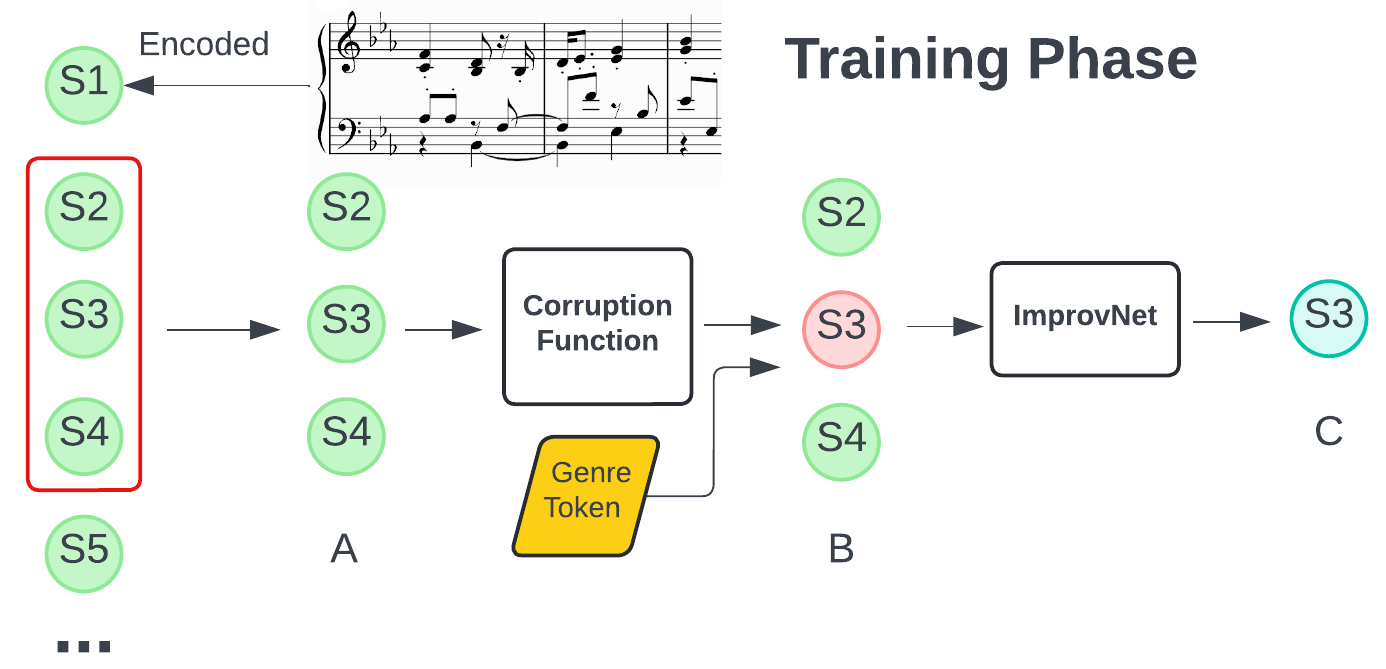}
    \caption{A) A random segment from $S$ is selected with context segments around it. B) This segment is corrupted with a corruption function. C) ImprovNet refines it back.}
    \label{fig:training-phase}
\end{figure}

\begin{figure}[h!]
\centering
\includegraphics[width=8cm]{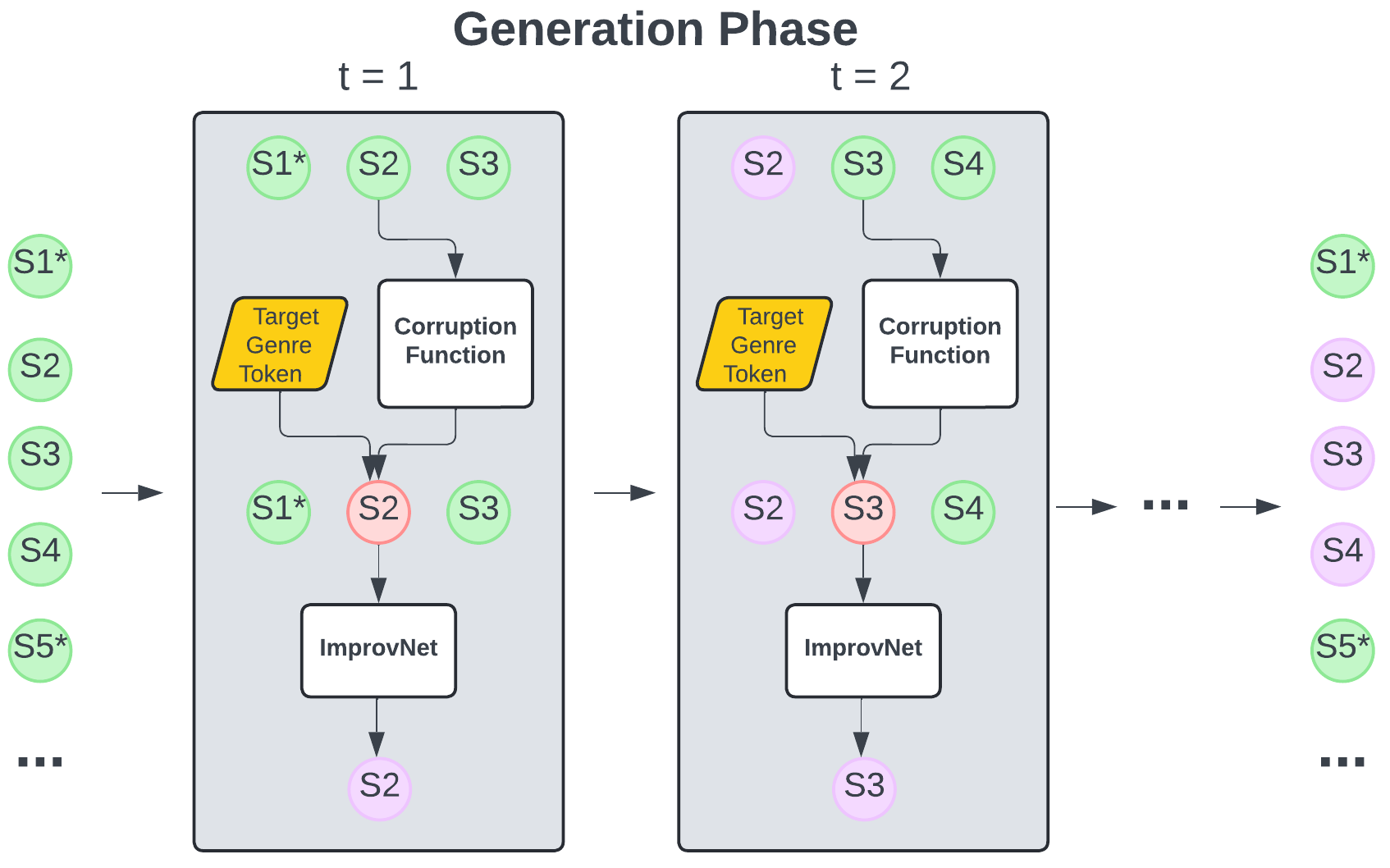}
\caption{A single pass iterates over all the segments in $S$. Segments marked with * are preserved. The other segments are corrupted with the corruption function and refined back.}
\label{fig:generation-phase}
\end{figure}

\subsection{ImprovNet Overview}
ImprovNet relies on a corruption-refinement training strategy similar to \cite{bhandari2025yin}, to generate music in an improvisational style of the original composition. The entire piece is segmented into 5-second segments as seen in Fig.~\ref{fig:aria-tokenizer}.

During training (Fig. ~\ref{fig:training-phase}), a random segment is selected with context segments on its left and right. This segment is corrupted using one of the corruption functions described in Section \ref{corruption-functions}. The corrupted segment, combined with the original context, is fed into a transformer encoder along with conditional tokens for the genre and corruption type. The transformer decoder learns to reconstruct the original segment.

During generation (Fig. ~\ref{fig:generation-phase}), 5-second segments are processed iteratively. A chosen segment is corrupted and refined based on a probability similar to that of training. For cross-genre improvisation, refinement conditions on the target genre token. This process iterates from beginning to end several times, gradually shaping the piece into the target genre.



\subsection{Formulation of Refinement Framework}

Let $S$ denote all 5-second segments separated by the $\langle T \rangle$ tokens. We can represent the encoding for a complete musical sequence as: 
\begin{equation}
    S=(s_1, \dots, s_N)
    \label{eq:musical-sequence}
\end{equation}
During training, we randomly sample a segment $s_i$ for $i < N$ from the sequence. To construct the input $S_{input}$, we include $(s_{i-L}, \ldots, s_{i-1})$ and $(s_{i+1}, \ldots, s_{i+R})$ for the $L$ and $R$ segments of context around $s_i$ where $L$ and $R$ represent the lengths of the left context and right context, respectively. This gives us the input $S_{input}$ as follows:
\begin{equation}
    S_{input} = (s_{i-L}, \ldots, s_{i-1}, s_{i}, s_{i+1}, \ldots, s_{i+R})
    \label{eq:input-sequence}
\end{equation}
When segments prior or after $s_i$ are unavailable such as the beginnings and endings of the sequence, we simply remove them from $S_{input}$. We leave the left and right context segments as they are, but corrupt $s_i$ with a randomly chosen corruption function from $f_{j\in J}$. The list of corruption functions is described in Section \ref{corruption-functions}. This process produces the corrupted segment \( s_{corrupted} = f_j(s_i) \), which is then transformed by prepending a genre label token \( \langle G \rangle \), resulting in the corrupted segment \( s_{c,g} \) constructed as follows:
\begin{equation}
    s_{c,g} = (\langle \texttt{sep} \rangle, \langle G \rangle, s_{corrupted}, \langle \texttt{sep} \rangle),
    \label{eq:corruption-fn}
\end{equation}
The corrupted segment \( s_c \) is prepended with a genre label token \( \langle G \rangle \), resulting in the creation of the following new corrupted input sequence:
\begin{equation}
    S_{c, g} = (s_{i-L}, \ldots, s_{i-1}, s_{c, g}, s_{i+1}, \ldots, s_{i+R})
    \label{eq:corrupt-sequence}
\end{equation}
During training, an encoder-decoder transformer \cite{vaswani2017attention} model $r_\theta$ is trained on the ground truth to refine back the corrupted segment. The inputs to the transformer encoder are the corrupted input sequence from equation \ref{eq:corrupt-sequence}, a conditional genre token $C_o$ based on the original genre of the musical sequence, and $j$, the type of corruption function selected. The auto-regressively refined output $S_r$ can be defined as:
\begin{equation}
S_r = r_\theta(S_{c,g})
\label{eq:refined-sequence}
\end{equation}

During generation, we iterate on each segment $s_i$ for $i \in \{1, 2, \ldots, N\}$ of the complete musical sequence $S$ as Equation \ref{eq:musical-sequence}. At every iteration, $s_i$ is corrupted with a user-defined or randomly chosen corruption function $f_j$ just as in Equation \ref{eq:corruption-fn} and refined back as Equation \ref{eq:refined-sequence}. The refinement at each iteration considers the left and right context segments just as in Equation \ref{eq:corrupt-sequence} but with a target genre conditional token $C_t$ instead of $C_o$ along with a corruption function token $j$ for Equation \ref{eq:refined-sequence}. When all segments have been iterated in the musical sequence, we refer to this as a single pass $P$. There could be multiple passes $P_q$ such that $q \in \{1, \dots, Q\}$. Each pass is also subject to a corruption rate $\alpha$ where the segment $s_i$ within each pass $P_q$ would be corrupted and refined back if $P_q<\alpha$ or left untouched otherwise. Furthermore, $W_i$ with $i \in \{1, \ldots, N\}$ is an indicator matrix that specifies whether corruption is applied to segments, where 0 indicates that there is no corruption and 1 indicates corruption. The refined output \( S_{r(q)} \) after passing through the \( q \)-th pass can be formulated as:
\begin{equation}
\begin{aligned}
    &\text{For } q = 1, \dots, Q, \\
    &\quad \text{For } i = 1, \dots, N, \\
    &s_{r(q)} = 
    \begin{cases}
        r_\theta(s_{r(q-1)}) & \text{if } (P_q < \alpha) \text{ and } W_i = 1, \\
        s_{r(q-1)} & \text{otherwise}.
    \end{cases}
\end{aligned}
\label{eq:multiple-passes}
\end{equation}

During training, the context lengths of $L$ and $R$ are randomly chosen between 1 and 5. During generation, we keep the first and last segments $s_1$ and $s_N$ intact. Furthermore, the list of segments $W$ that are preserved is determined based on a self-similarity and novelty algorithm from \cite{SSMNet}. ImprovNet users can decide to retain more or less original musical segments based on a preservation ratio between 0 to 1 depending on their preference. For our objective experiments, we keep this value to 0.05, indicating that 5 out of 100 original segments $s_i$ would be retained at various times, while the rest would be transformed by Equation \ref{eq:multiple-passes}. The aim is to retain the novel segments, including the start of new musical sections, so that users can better relate to the original musical structure.

\subsection{Corruption Functions}
\label{corruption-functions}
During training, ImprovNet uses nine different corruption functions ${f_j}$ to refine sequences relative to ground truth, conditioned on genre tokens.

\subsubsection{Pitch Velocity Mask}
The pitch-velocity tokens within a $s_i$ segment are masked while the duration and onset tokens are kept intact. 

\subsubsection{Onset Duration Mask}
The onset and duration tokens are masked out. The goal is to add syncopation during the classical-to-jazz cross-genre improvisation task.

\subsubsection{Whole Mask}
The entire $s_i$ segment is masked and replaced with a special ``whole mask'' token. The model learns to generate all the notes omitted for this segment with respect to the context surrounding it. During training, when the whole mask corruption is randomly sampled, we further drop the right context segments $(s_{i+1}, \ldots, s_{i+R})$ 50\% of the time for the short prompt continuation and short infilling tasks in section \ref{short-prompt-continuation} and \ref{short-infilling}.

\subsubsection{Permute Pitch}
We shuffle the MIDI pitches of all notes within a $s_i$ segment while preserving their velocity values. 


\subsubsection{Permute Pitch Velocity}
We shuffle both the pitch and velocity MIDI values for all notes within $s_i$. 

\subsubsection{Fragmentation}
Fragmentation retains a subset of notes within $s_i$, randomly keeping 20\%–50\% of the notes at the beginning. The model learns to generate missing notes to produce useful variations of the original segment. 

\subsubsection{Incorrect Transposition}
We randomly shift 50\% of the notes in $s_i$ by $\pm$ 5 semitones while preserving their velocity, training the model to regenerate the correct pitches. This incorrect transposition, when conditioned on the jazz target genre, introduces jazzy melodies and harmonies, often transforming rapid semitone passages into chromatic scales.


\subsubsection{Note Modification}
Note modification randomly omits 10\%–40\% of notes spaced at least 50 milliseconds apart, adding their duration to the preceding note. New notes are then inserted after notes longer than 500 milliseconds, with a probability of 10\% - 40\%. The pitches of these new notes are chosen within $\pm$ 5 semitones, and their velocities are randomly selected between 45 and 105 in increments of 15.

\subsubsection{Skyline}
We use the Skyline \cite{uitdenbogerd1999melodic} method for the extraction of the melody by selecting the highest pitch values. Notes within 50 milliseconds are treated as chords to account for slight timing offsets in human performance, and their MIDI velocity is fixed at 90. Although not always melodic \cite{kosta2022deep}, Skyline serves as a corruption function to train ImprovNet to regenerate omitted notes, allowing harmonization with a monophonic melody, as discussed in Section \ref{Harmonization}, while learning dynamics for more expressive music.

\subsection{Self-Supervised Pre-Training and Fine-Tuning}
ImprovNet uses a self-supervised corruption refinement strategy, where segments are corrupted with predefined functions to help the model learn musical structures and relationships. This enables learning from weak genre labels without requiring richly annotated datasets, capturing genre-specific traits. ImprovNet was pre-trained on classical music and fine-tuned on both classical and jazz datasets, addressing the limited availability of polyphonic solo jazz piano data. To improve accuracy, some datasets were retranscribed from audio to MIDI using a state-of-the-art transcription model\footnote{https://github.com/EleutherAI/aria-amt}. More details on this process are provided in Section \ref{datasets}.

\subsection{Iterative Generation}
In this section, we describe the tasks ImprovNet can perform through its iterative generation framework.

\subsubsection{Style-aware Improvisations}
By conditioning ImprovNet on a genre label, users can generate two types of improvisations. The first is {\bfseries cross-genre} improvisation (CGI), where a genre label different from the original is used as conditioning. The second is {\bfseries intra-genre} improvisation (IGI), where the genre label remains the same, preserving the original genre.

Users can also select different corruption functions with corresponding corruption rates ($\alpha$) for each pass and adjust the context window size during refinement. Initial experiments show that CGI works well with smaller right context sizes ($R$, e.g., 1-2) or additional passes using progressively lower corruption rates compared to IGI. However, the optimal setup varies depending on the unique characteristics of each piece.

\subsubsection{Short Prompt Continuation}
\label{short-prompt-continuation}
ImprovNet can generate short continuations (5-20 seconds) of a prompt using whole mask corruption while trimming the right context during the refinement process. For example, given a 20-second prompt (4 segments) at the start of a piece and a task to generate the next 15 seconds (3 segments), we adjust Equation \ref{eq:multiple-passes} as follows:
\begin{multline}
    \text{For } q = 1, \text{ let } P_q: \forall i \in \{5, 6, 7\}, \\
    \text{ compute } s_i \text{ as } S_{r} = r_\theta(S_{c,g}[:i]),
    \label{eq:short-prompt}
\end{multline}
where $S_{c,g}[\colon i] = (s_{i-L}, \ldots, s_{i-1}, s_{c,g})$. Here, the right context is trimmed to the index $i$ to iteratively refine the sequence. The corrupted segment $s_{c,g}$ is generated using the whole-mask corruption method for the specific corruption function $f_{j=3}$ described in Equation \ref{eq:corruption-fn}. We find the coherency of the generated music starts deteriorating beyond 20 seconds, as the model isn't explicitly trained for long continuations.

\subsubsection{Short Infilling}
\label{short-infilling}
ImprovNet can also perform short infilling (5–20 seconds) by considering both left and right contexts. For example, consider the scenario where the first 20 seconds and 40–60 seconds are provided as the left and right side contexts and 20–40 seconds need to be infilled by ImprovNet. The infilling process begins by generating segments $s_5$, $s_6$ and $s_7$ (20–35 seconds) as a continuation of the left context using the whole mask corruption. During this process, the right side context is not provided to ImprovNet as described in equation \ref{eq:short-prompt}. For the final segment $s_8$ (35–40 seconds), the left and right contexts are included in the generation, as shown in Equation \ref{eq:multiple-passes} to ensure a smooth transition to the segments after $s_8$.

\subsubsection{Harmonization}\label{Harmonization}
The skyline algorithm used as a corruption function helps ImprovNet learn to generate notes below a MIDI pitch. During generation, when a monophonic melody is provided, we use the skyline corruption function for Equation \ref{eq:multiple-passes} along with constraints to the model's logits for the first pass to harmonize the entire melody. Specifically, we constrain the logits during autoregressive sampling to produce a chord beneath the first melody note of each segment \( s_i \) in the musical sequence \( S \). The standard token generation process follows the probability distribution:
\begin{equation}
P(x_t \mid x_{<t}) = \frac{\exp\left(\frac{z_t}{\tau}\right)}{\sum_k \exp\left(\frac{z_k}{\tau}\right)},
\label{eq:token-generation}
\end{equation}
where \( z_t \) are the logits for token \( x_t \), \( \tau \) is the temperature controlling the sharpness of the distribution, and \( x_{<t} \) denotes the preceding sequence of tokens. To enforce chord generation, we use a logit constraint for the next \( N = 3 \) onset tokens, where \( N \) is user-defined. Specifically, for each onset token \( x_t \), the logits \( z_t \) are set to \(-\infty\) for all tokens outside the onset range \([0, 50 \, \text{ms}]\) relative to the first melody note \( \text{onset}(x_1) \). This is to account for slight chord onset deviations in human performance. Mathematically, the constrained logits are:
\begin{equation}
z_t = 
\begin{cases} 
z_t & \text{if } 0 \leq \text{onset}(x_t) - \text{onset}(x_1) \leq 50 \, \text{ms}, \\
-\infty & \text{otherwise}.
\end{cases}
\label{eq:logit-constraint}
\end{equation}

The encoding for a note within segment $s_i$ consists of three tokens (see Fig.~\ref{fig:aria-tokenizer}): Onset, Duration, and Pitch-Velocity tokens. While the logits for the onset tokens are constrained as described in equation \ref{eq:logit-constraint}, the note duration and pitch-velocity tokens are generated with the standard autoregressive process. After the first chord is generated, all subsequent onset, duration, and pitch-velocity tokens within segment \( s_i \) are generated with the standard autoregressive approach, allowing the model to follow its learnt stylistic patterns.

The constraints on the first note of the segment \( s_i \) force the model to harmonize the other melody notes within the segment, as the subsequent tokens naturally align with the first generated chord. Without these constraints, there is no inherent pressure for the model to produce chords as the right context segments are monophonic. After the first pass, we run a few more passes using the skyline corruption function but without constraints, allowing the generation to normalize and smoothen over iterations. Most importantly, by conditioning on a different target genre during generation, we can achieve harmonic genre style conversion, such as generating jazz chords for a folk, pop, or classical melody. We show the effectiveness of this approach in Section \ref{results}.

\section{Experiments}

\subsection{Datasets}
\label{datasets}
We pre-train ImprovNet on a retranscribed version of the ATEPP dataset \cite{zhang2022atepp}, which comprises $\sim$1000 hours of classical solo piano music. For fine-tuning, we use three datasets: Maestro \cite{hawthorne2018enabling} with 177 hours of classical piano music, PiJAMA \cite{edwards2023pijama} with over 200 hours of jazz piano music, and the Doug McKenzie dataset\footnote{https://bushgrafts.com/midi/}, which includes 307 short jazz MIDI pieces. Maestro and Doug McKenzie contain accurate expressive renditions, while PiJAMA was retranscribed to improve the MIDI accuracy of the original audio. Some jazz MIDI files with multiple instruments were processed to retain only piano tracks. We used the entire dataset for pretraining but randomly extracted 100 MIDI files at the fine-tuning stage as a test set for evaluation. Furthermore, we used 100 pieces taken at random from the Wikifonia dataset \cite{simonetta2018symbolic} to evaluate the harmonization task.

\subsection{Training Details}
\subsubsection{Pretraining and Fine-tuning}
ImprovNet uses an encoder-decoder transformer architecture for pre-training and fine-tuning. The encoder consists of 12 layers with 8 attention heads, a hidden size of 512, and an intermediate feedforward size of 2048, while the decoder has the same configuration. The maximum sequence lengths for the encoder and decoder are set to 2048 and 512 tokens, respectively, ensuring that the model can sufficiently process a maximum of 11 segments (55 seconds) of Aria's tokenized encoding as input during training. During pre-training, the model was trained for 360K steps with a batch size of 4, a learning rate of \(1 \times 10^{-4}\), a weight decay of 0.01, and a warm-up ratio of 0.3. Fine-tuning was performed for 318K steps, with the learning rate reduced to \(5 \times 10^{-5}\), a weight decay of 0.01, and a warm-up ratio of 0.1. The batch size remained at 4, but the gradient accumulation steps were increased to 3 to handle the smaller size of the dataset during fine-tuning and to stabilize the updates.

\subsubsection{Genre Classifier}
\label{genre-classifier}
To evaluate the effectiveness of the CGI task, a symbolic genre classification model was trained as a secondary task. During training, the classifier processes three consecutive $s_i$ segments randomly selected from the sequence $S$. During generation, probabilities are averaged by applying a sliding window to the entire sequence $S$, considering three segments at a time. The classifier model shares the same encoder structure as the primary model, with a maximum sequence length of 1024 tokens but with a binary cross-entropy learning objective. Training was carried out in 16K steps with a batch size of 64, a learning rate of \(1 \times 10^{-4}\), and a weight decay of 0.01. Dropout was applied to the encoder with a probability of 0.1 to prevent overfitting.

\section{Evaluations}

\subsection{Objective Evaluation}
\subsubsection{Style-aware Improvisation}
Evaluating improvisations is a challenging task. We used two objective evaluation metrics to understand the impact of corruption functions, corruption rates, and number of passes on genre conversion and structural similarity with respect to the original composition. 

\begin{itemize}
    \item Genre Classifier: We train a genre classifier (see Section \ref{genre-classifier}) to predict the probability of the genre of the original and generated compositions.
    \item Structural Similarity Score: We derive a chroma-based structural similarity matrix from the audio of the original and generated compositions and compute the Pearson correlation between the two matrices. This indicates how similar the generation is to the original composition.
\end{itemize}

In the classical to jazz CGI task, we expect the jazz probability to increase depending on the corruption function, the corruption rate, and the number of passes specified.

\subsubsection{Short Prompt Continuation and Short Infilling}
To evaluate ImprovNet for short prompt continuation and short infilling tasks, we compare the generated outputs with Anticipatory Music Transformer \cite{thickstun2023anticipatory} (AMT), a state-of-the-art music infilling and continuation model. For short prompt continuation, we use both models to generate 10 seconds of music to follow a 20 second prompt from the test data. Similarly, for short infilling, we provide both models 0:20 and 30:50 seconds as context and generate 20:30 seconds. We compare these two models with the ground truth in the following metrics.
\begin{itemize}
    \item Pitch Class KL Divergence: KL-Divergence computed over a 12-d vector representing the frequency of each pitch class between the original and generated notes.
    \item Pitch Class Transition Matrix (PCTM)~\cite{yang2020eval}: A 12×12 matrix showing the frequency of pitch class transitions, with cosine similarity used to assess differences in interval preferences and inter-note variation.
    \item Note Density: The average number of notes played in 5-second segments of a piece of music.
    \item Average Inter-Onset-Interval (IOI): The average time gap (in seconds) between two consecutive notes \cite{yang2020eval}.
    \item Unique Pitches: The number of unique MIDI pitches generated, indicating pitch diversity.
\end{itemize}

\subsubsection{Harmonization}
To the best of our knowledge, ImprovNet is the first model to generate expressive genre-conditioned harmonies for a melody, altering the dynamics of the original melody as well. Given the importance of logit constraints for this task, we evaluated the constrained and unconstrained versions of ImprovNet, along with randomly generated chords and ground truth. The objective evaluation uses the following metrics averaged over generations:

\begin{itemize}
    \item Polyphony Rate: Ratio of the number of time steps in which multiple pitches are played to the total time steps. 
    \item Note Density: Average number of notes played in 5-second segments of a music piece.
    \item Tonal Tension Diameter: The largest distance between two notes in a chord. 
    \item Chord Diversity: Number of unique chords in a piece.
    \item Pitch in Scale Rate: Percentage of time the notes adhere to the musical scale of the pieces.
\end{itemize}

\subsection{Subjective Evaluation}
We conducted a subjective assessment of ImprovNet using a Qualtrics listening test with 28 participants, who spent approximately 40 minutes providing quantitative and qualitative feedback on the generated examples. 18 participants had 3+ years of formal training in music theory (9 with 10+ years), and 20 had 3+ years of formal training on a musical instrument including voice (11 with 10+ years). 20 and 18 participants were familiar with jazz and classical genres, respectively.

As this work is the first expressive performance generation for genre-style-based improvisation and harmony generation tasks, we used the ground truth as the baseline to assess our model's performance. Complete music examples around 2-4 minutes were shown for these tasks. For the short prompt continuation task, we compared our model with AMT \cite{thickstun2023anticipatory}. Participants were randomly assigned to one of four versions of the survey, each featuring different music examples.

Participants rated three music examples in the first section. The first piece, a classical genre example, was disclosed as the ground truth. The second and third examples were generated in either a classical or jazz improvisational style, with their genre style undisclosed. Participants were asked to identify the jazz improvisation and also to make notes on the musical theme and structure, including theme repetitions and variations, and new section starts while listening to the examples. They then rated the generated examples on the following questions:
\begin{itemize}
    \item Interestingness: How interesting does the music sound to you on a scale of 1 (not at all) to 5 (very interesting)?
    \item Human-like: How human-like does the music sound to you on a scale of 1 (not at all) to 5 (very human-like)?
    \item Overall: How much do you like the music overall on a scale of 1 (not at all) to 5 (a lot)?
    \item Structural Similarity: To what extent does the generated music example refer to the original music example on a scale of 1 (not at all) to 5 (a lot)? Consider repetition and variations of the original theme and the sequence of musical sections in your answer.
\end{itemize}

In Section 2, participants were shown two examples: the original composition and a generated variation in the same genre style. Although the questions mirrored those of Section 1, participants were asked to take notes on the musical theme, melody, harmony, rhythm, and style.

Section 3, the harmonization task, followed the same format as in Section 1. Participants were presented with an original melody and harmony from the Wikifonia dataset, along with classical and jazz harmonizations of the melody. Both ImprovNet generations used logit constraints, as explained in Section \ref{Harmonization}. Participants identified the jazz harmonization and answered questions on how well the harmony fit the melody (Match), how interesting the harmonic progressions were, and how much they liked the harmony overall. 


In Section 4 on short prompt continuation, we presented participants with 2 generation excerpts of 20 seconds on an initial prompt of 20 seconds from ImprovNet and the AMT baseline. Following the evaluation in \cite{thickstun2023anticipatory}, we asked participants to rate which generation they found more conventionally musical, while also presenting them with the option to select both as equally musical.

\section{Results}
\label{results}

\subsection{Objective Results}

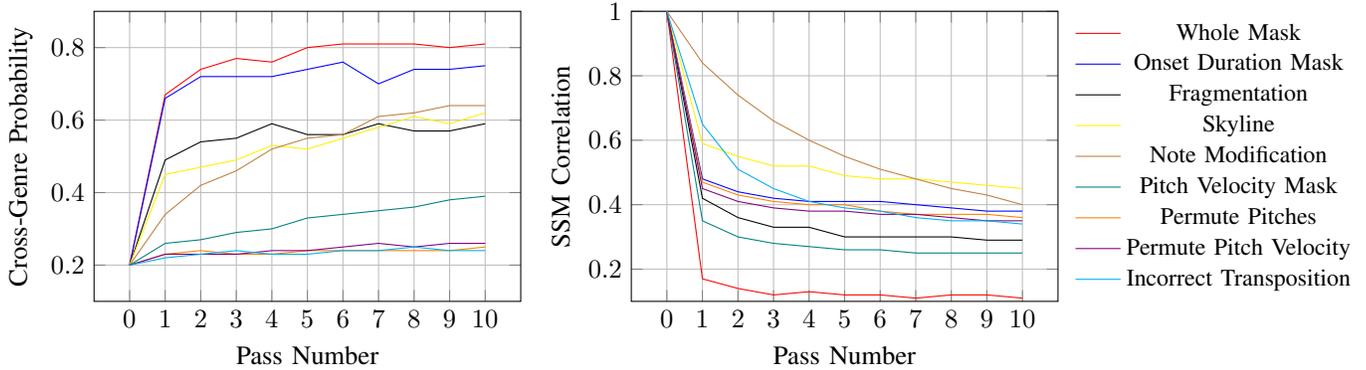
\begin{figure*}[h!]
\centering
\makebox[\textwidth][l]{
    \begin{subfigure}[b]{0.40\textwidth}
        \centering
        \begin{tikzpicture}
        \begin{axis}[
            width=\textwidth,
            height=0.75\textwidth,
            xlabel={Pass Number},
            ylabel={Cross-Genre Probability},
            grid=major,
            xtick=data,
            ymin=0.1, ymax=0.9,
            cycle list name=color list
        ]
        \addplot coordinates {(0,0.2) (1,0.67) (2,0.74) (3,0.77) (4,0.76) (5,0.80) (6,0.81) (7,0.81) (8,0.81) (9,0.80) (10,0.81)};
        \addplot coordinates {(0,0.2) (1,0.66) (2,0.72) (3,0.72) (4,0.72) (5,0.74) (6,0.76) (7,0.70) (8,0.74) (9,0.74) (10,0.75)};
        \addplot coordinates {(0,0.2) (1,0.49) (2,0.54) (3,0.55) (4,0.59) (5,0.56) (6,0.56) (7,0.59) (8,0.57) (9,0.57) (10,0.59)};
        \addplot coordinates {(0,0.2) (1,0.45) (2,0.47) (3,0.49) (4,0.53) (5,0.52) (6,0.55) (7,0.58) (8,0.61) (9,0.59) (10,0.62)};
        \addplot coordinates {(0,0.2) (1,0.34) (2,0.42) (3,0.46) (4,0.52) (5,0.55) (6,0.56) (7,0.61) (8,0.62) (9,0.64) (10,0.64)};
        \addplot coordinates {(0,0.2) (1,0.26) (2,0.27) (3,0.29) (4,0.30) (5,0.33) (6,0.34) (7,0.35) (8,0.36) (9,0.38) (10,0.39)};
        \addplot coordinates {(0,0.2) (1,0.23) (2,0.24) (3,0.23) (4,0.23) (5,0.24) (6,0.24) (7,0.24) (8,0.24) (9,0.24) (10,0.25)};
        \addplot coordinates {(0,0.2) (1,0.23) (2,0.23) (3,0.23) (4,0.24) (5,0.24) (6,0.25) (7,0.26) (8,0.25) (9,0.26) (10,0.26)};
        \addplot coordinates {(0,0.2) (1,0.22) (2,0.23) (3,0.24) (4,0.23) (5,0.23) (6,0.24) (7,0.24) (8,0.25) (9,0.24) (10,0.24)};
        \end{axis}
        \end{tikzpicture}
        \caption{Effect of multiple passes on cross-genre probability.}
        \label{fig:obj_improvisations_genre}
    \end{subfigure}
    \begin{subfigure}[b]{0.40\textwidth}
        \centering
        \begin{tikzpicture}
        \begin{axis}[
            width=\textwidth,
            height=0.75\textwidth,
            xlabel={Pass Number},
            ylabel={SSM Correlation},
            grid=major,
            xtick=data,
            ymin=0.1, ymax=1.0,
            legend style={font=\small, at={(1.02,0.5)}, anchor=west, draw=none},
            cycle list name=color list
        ]
        \addplot coordinates {(0,1.0) (1,0.17) (2,0.14) (3,0.12) (4,0.13) (5,0.12) (6,0.12) (7,0.11) (8,0.12) (9,0.12) (10,0.11)};
        \addlegendentry{Whole Mask}
        \addplot coordinates {(0,1.0) (1,0.48) (2,0.44) (3,0.42) (4,0.41) (5,0.41) (6,0.41) (7,0.40) (8,0.39) (9,0.38) (10,0.38)};
        \addlegendentry{Onset Duration Mask}
        \addplot coordinates {(0,1.0) (1,0.42) (2,0.36) (3,0.33) (4,0.33) (5,0.30) (6,0.30) (7,0.30) (8,0.30) (9,0.29) (10,0.29)};
        \addlegendentry{Fragmentation}
        \addplot coordinates {(0,1.0) (1,0.59) (2,0.55) (3,0.52) (4,0.52) (5,0.49) (6,0.48) (7,0.48) (8,0.47) (9,0.46) (10,0.45)};
        \addlegendentry{Skyline}
        \addplot coordinates {(0,1.0) (1,0.84) (2,0.74) (3,0.66) (4,0.60) (5,0.55) (6,0.51) (7,0.48) (8,0.45) (9,0.43) (10,0.40)};
        \addlegendentry{Note Modification}
        \addplot coordinates {(0,1.0) (1,0.35) (2,0.30) (3,0.28) (4,0.27) (5,0.26) (6,0.26) (7,0.25) (8,0.25) (9,0.25) (10,0.25)};
        \addlegendentry{Pitch Velocity Mask}
        \addplot coordinates {(0,1.0) (1,0.47) (2,0.43) (3,0.41) (4,0.40) (5,0.40) (6,0.38) (7,0.37) (8,0.37) (9,0.37) (10,0.36)};
        \addlegendentry{Permute Pitches}
        \addplot coordinates {(0,1.0) (1,0.45) (2,0.41) (3,0.39) (4,0.38) (5,0.38) (6,0.37) (7,0.37) (8,0.36) (9,0.35) (10,0.35)};
        \addlegendentry{Permute Pitch Velocity}
        \addplot coordinates {(0,1.0) (1,0.65) (2,0.51) (3,0.45) (4,0.41) (5,0.39) (6,0.38) (7,0.36) (8,0.35) (9,0.35) (10,0.34)};
        \addlegendentry{Incorrect Transposition}
        \end{axis}
        \end{tikzpicture}
        \caption{Effect of multiple passes on structural similarity.}
        \label{fig:obj_improvisations_ssm}
    \end{subfigure}
}
\caption{Cross-genre improvisations generated with all corruption functions using a corruption rate of 1.0.}
\end{figure*}

\begin{singlespace}
\begin{figure}[h!]
\centering
\begin{tikzpicture}
    \begin{axis}[
        name=plot1,
        xlabel={Corruption Rate},
        ylabel={Cross-Genre Probability},
        xmin=0.25, xmax=1.0,
        ymin=0.2, ymax=0.6,
        xtick={0.25, 0.5, 0.75, 1.0},
        ytick={0.3, 0.4, 0.5, 0.6},
        legend pos=south west,
        legend style={font=\small, at={(0.02,0.0)}, anchor=south west, draw=none},
        grid=major, 
        grid style={solid, gray!50},
        cycle list name=color list, 
        width=0.32\textwidth,
        height=0.28\textwidth,
    ]
    \addplot[
        color=blue,
        thick,
        mark=o,
        mark options={solid},
    ] coordinates {
        (0.25, 0.42)
        (0.5, 0.47)
        (0.75, 0.51)
        (1.0, 0.53)
    };
    \addlegendentry{Cross-Genre Probability}
    \end{axis}
    
    \begin{axis}[
        name=plot2,
        xlabel={Corruption Rate},
        ylabel={SSM Correlation},
        xmin=0.25, xmax=1.0,
        ymin=0.2, ymax=0.6,
        xtick=\empty,
        ytick={0.3, 0.4, 0.5, 0.6},
        axis x line=none,
        axis y line=right,
        legend pos=south west,
        legend style={font=\small, at={(0.02,0.12)}, anchor=south west, draw=none},
        width=0.32\textwidth,
        height=0.28\textwidth,
    ]
    \addplot[
        color=red,
        thick,
        mark=square,
        mark options={solid},
    ] coordinates {
        (0.25, 0.44)
        (0.5, 0.36)
        (0.75, 0.33)
        (1.0, 0.31)
    };
    \addlegendentry{SSM Correlation}
    \end{axis}
\end{tikzpicture}
\caption{Effect of various corruption rates on cross-genre probability and structural similarity.}
\label{fig:corruption_rate}
\end{figure}
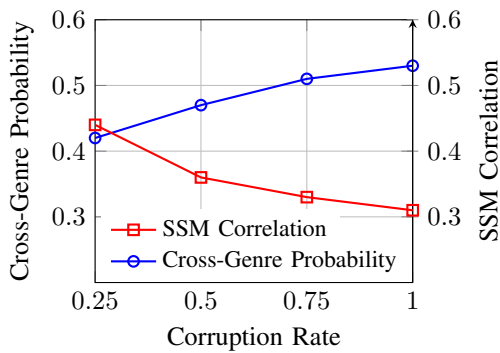
\end{singlespace}

\begin{table}[h!]
\centering
\small
\setlength{\tabcolsep}{3pt} 
\renewcommand{\arraystretch}{1.1} 
\begin{tabular}{lccccc}
\hline
\textbf{Model} & \makecell{\textbf{Avg.}\\\textbf{IOI}} & \makecell{\textbf{Note}\\\textbf{Density}} & \makecell{\textbf{Unique}\\\textbf{Pitches}} & \makecell{\textbf{PCTM}\\\textbf{Cosine Sim}$\uparrow$} & \makecell{\textbf{Pitch}\\\textbf{Class KL}$\downarrow$} \\
\hline
AMT       & \textbf{0.1386} & 67.66 & 25.15 & 0.3074 & 1.6084 \\
ImprovNet & 0.1244 & \textbf{37.49} & \textbf{28.53} & \textbf{0.3470} & \textbf{1.2500} \\
\hline
Original & 0.1405 & 30.85 & 27.07 & -     & -     \\
\hline
\end{tabular}
\caption{Short prompt continuation for AMT, ImprovNet, and Original. IOI = Inter-onset interval, PCTM = Pitch class transition matrix.}
\label{tab:obj_short_continuation}
\end{table}

\begin{table}[h!]
\centering
\small
\setlength{\tabcolsep}{3pt} 
\renewcommand{\arraystretch}{1.1} 
\begin{tabular}{lccccc}
\hline
\textbf{Model} & \makecell{\textbf{Avg.}\\\textbf{IOI}} & \makecell{\textbf{Note}\\\textbf{Density}} & \makecell{\textbf{Unique}\\\textbf{Pitches}} & \makecell{\textbf{PCTM}\\\textbf{Cosine Sim}$\uparrow$} & \makecell{\textbf{Pitch}\\\textbf{Class KL}$\downarrow$} \\
\hline
AMT       & \textbf{0.1508} & 49.63 & 25.45 & 0.3624 & 1.4593 \\
ImprovNet & 0.1224 & \textbf{36.30} & \textbf{28.78} & \textbf{0.4036} & \textbf{0.8907} \\
\hline
Original & 0.1394 & 32.44 & 27.88 & -     & -     \\
\hline
\end{tabular}
\caption{Short infilling for AMT, ImprovNet, and Original.}
\label{tab:obj_short_infilling}
\end{table}

\begin{table}[h!]
\centering
\small
\setlength{\tabcolsep}{3pt} 
\renewcommand{\arraystretch}{1.1} 
\begin{tabular}{lccccc}
\hline
\textbf{Model} & \makecell{\textbf{Poly.}\\\textbf{Rate}} & \makecell{\textbf{Note}\\\textbf{Density}} & \makecell{\textbf{Tonal}\\\textbf{Tension}\\\textbf{Diameter}} & \makecell{\textbf{Chord}\\\textbf{Diversity}} & \makecell{\textbf{Pitch in}\\\textbf{Scale Rate}} \\
\hline
ImprovNet wc       & 0.91 & 35.95 & 0.62 & 18.30 & 80.17 \\
ImprovNet wo/c & 0.25 & 11.06 & 0.75 & 13.62 & 80.26 \\
Random Chords & - & - & 0.91 & 37.38 & 63.64 \\
Monophonic & 0.00 & 8.71 & - & - & - \\
\hline
Original & 0.96 & 19.20 & 0.46 & 13.29 & 79.40 \\
\hline
\end{tabular}
\caption{ImprovNet harmonization with and without constraints, random chords, monophonic melody and original.}
\label{tab:obj_harmony}
\end{table}

For the CGI task in Fig.~\ref{fig:obj_improvisations_genre}, each corrupted function refined over multiple passes increases the cross-genre (classical to jazz) probability score. The whole mask and onset-duration mask corruption functions have the highest increases in probability, while the incorrect transposition and permutation functions have the lowest. We also see the sharpest decrease in the SSM correlation score in Fig.~\ref{fig:obj_improvisations_ssm} for the whole mask function, indicating that the generated pieces deviate significantly in structure from the original composition. It is interesting to see that the note modification and skyline functions retain more of the original music structure over multiple passes than other functions do. However, we note that the corruption rate here was set to a maximum of 1.0. As seen in Fig.~\ref{fig:corruption_rate}, the changes in cross-genre probability and SSM correlation scores are more gradual when the corruption rate is lower. Thus, this shows that the user can control the improvisation target genre and structural similarity based on different corruption functions and corruption rates over multiple passes.

Tables \ref{tab:obj_short_continuation} and \ref{tab:obj_short_infilling} show ImprovNet's superior performance over the AMT baseline in adhering closer to the original distribution for the short continuation and short infilling tasks. We attribute this success to the various corruptions ImprovNet sees during training, which serves as a form of data augmentation for our model and improves its internal representation of music. Table \ref{tab:obj_harmony} shows the effectiveness of ImprovNet with constraints for harmonizing monophonic melodies, as seen in the polyphonic rate. ImprovNet without constraints is not able to harmonize melodies as effectively and very frequently returns the unharmonized melody. This is because there is no inherent pressure on the model as the context around it is monophonic. ImprovNet with constraints is able to produce a more diverse range of chords as seen in the chord diversity metric. The higher note density implies dense chords that contain many voices. Although the generated samples show a relatively higher harmonic tension compared to the original, it is much lower than randomly generated chords and is also closer to the original in the pitch in the scale rate metric.

\subsection{Subjective Results}

\begin{table}[h!]
\centering
\small 
\setlength{\tabcolsep}{4pt} 
\renewcommand{\arraystretch}{1.1} 
\begin{tabular}{lccccc}
\hline
\textbf{Model} & \textbf{Interest} & \textbf{Human.} & \textbf{Overall} & \textbf{Struct.} & \textbf{Genre} \\
\hline
ImprovNet cgi & 3.36 & 2.71 & 3.11  & 3.21 & 0.79\\
ImprovNet igi & 3.39 & 3.25 & 3.21  & 4.07 & -\\
Original  & 3.43 & 5* & 3.64 & - & -\\
\hline
\end{tabular}
\caption{Cross-genre and intra-genre improvisations. \\
Human-like score is assumed as 5 for the original as it was disclosed due to the nature of questions asked.}
\label{tab:style_aware_improvisation}
\end{table}

\begin{table}[h!]
\centering
\small 
\setlength{\tabcolsep}{4pt} 
\renewcommand{\arraystretch}{1.1} 
\begin{tabular}{lccccc}
\hline
\textbf{Model} & \textbf{Interest} & \textbf{Human.} & \textbf{Overall} & \textbf{Struct.} \\
\hline
ImprovNet igi & 3.57  & 3.54  & 3.39  & 3.93  \\
Original  & 4.11  & 5*    & 4.07  & -     \\
\hline
\end{tabular}
\caption{Intra-genre improvisations. Human-like score is assumed as 5 for the original as it was disclosed due to the nature of questions asked.}
\label{tab:intra_genre_improvisation}
\end{table}

\begin{table}[h!]
\centering
\small 
\setlength{\tabcolsep}{4pt} 
\renewcommand{\arraystretch}{1.1} 
\begin{tabular}{lccccc}
\hline
\textbf{Model} & \textbf{Interest} & \textbf{Match} & \textbf{Overall} & \textbf{Genre} \\
\hline
ImprovNet cgh & 3.54      & 2.65      & 2.92      & 0.76      \\
ImprovNet igh & 2.58      & 3.27      & 2.54      & -      \\
Original  & 3.38      & 3.89      & 3.31      & -      \\
\hline
\end{tabular}
\caption{Cross-genre and intra-genre harmonization}
\label{tab:harmony_generation}
\end{table}

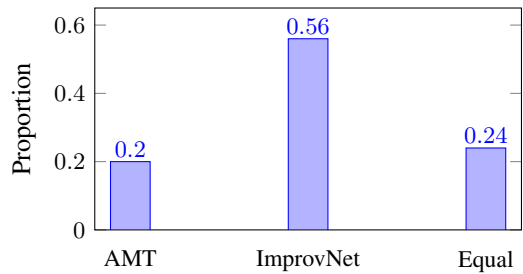
\begin{figure}[h!]
\centering
\begin{tikzpicture}
    \begin{axis}[
        width=0.4\textwidth, 
        height=0.25\textwidth, 
        ybar,
        bar width=15pt,
        symbolic x coords={AMT, ImprovNet, Equal},
        xtick=data,
        ymin=0, ymax=0.65, 
        ylabel={Proportion},
        nodes near coords,
        nodes near coords style={font=\small, anchor=south, inner sep=1.5pt},
        xtick style={draw=none},
        tick label style={font=\small},
        legend style={at={(0.5,-0.2)}, anchor=north, legend columns=-1},
    ]
        \addplot coordinates {(AMT,0.2) (ImprovNet,0.56) (Equal,0.24)};
    \end{axis}
\end{tikzpicture}
\caption{Musicality preference comparison among AMT, ImprovNet, and equal preference.}
\label{fig:short_prompt_continuation}
\end{figure}

Table \ref{tab:style_aware_improvisation} presents the results of Section 1 of the listening test on style-aware improvisations. It is encouraging to see that the generated samples are almost on par with the original in interestingness, and 79\% of the participants were able to identify the jazz-style conditioned example while giving a moderately high score for structural similarity. We conducted a one-sample binomial test for the genre identification task and received significant results ($p=0.0037$) for $\alpha=0.05$. It is not surprising to see IGI (classical to classical) rated higher for all metrics, as these are relatively closer in style to the original examples. We attribute the lower human-like score for CGI to occasionally distorted phrasal structures by ImprovNet, as some participants pointed out in the qualitative assessment.

In table \ref{tab:intra_genre_improvisation} referring to Section 2 (intra-genre improvisations) of the listening study, we see higher human-like scores than the CGI in Section 1 because the generated examples are closer to the original examples in style. However, there is a gap in the overall rating between the generated and original examples, which is expected to some extent due to the nature of improvisations compared to original works.

It is noteworthy to see ImprovNet cross-genre harmonizations score higher than the original examples in table \ref{tab:harmony_generation} for the interestingness of harmonic progressions. While we see a higher score for harmonic match with the melody for the intra-genre harmonizations, the chords generated often lacked diversity compared to their cross-genre counterpart, leading to lower scores for overall rating. Some participants reported higher dissonance levels in the cross-genre jazz harmonization examples, leading to lower scores for the harmony-melody match. We note that ImprovNet was not trained on the Wikifonia dataset, a key factor in the higher tonal tension. In addition, 76\% of the participants were able to correctly identify jazzy harmonies for melodies of the Wikifonia dataset. These results were also statistically significant ($p=0.0133$) for $\alpha=0.05$.

Finally, table \ref{fig:short_prompt_continuation} based on Section 4 (short prompt continuation) of the listening study shows that 56\% of the users preferred ImprovNet over AMT while 24\% equally preferred generations from both models. This correlates with Table \ref{tab:obj_short_continuation} of the objective evaluation and highlights ImprovNet's effectiveness in adhering to the given style of the short prompt.

\section{Discussion}


ImprovNet exhibits strong capabilities to generate expressive style-aware improvisations, while offering users the ability to control the intensity of style transfer and musical structure, as seen in the objective and subjective results. Additionally, its versatility extends to short prompt continuation and infilling tasks, which achieve better performance compared to the AMT baseline. Moreover, the skyline corruption algorithm combined with the constraints over the logits allows ImprovNet to harmonize a given melody with respect to the musical genre. 

For generating improvisations, there is no best corruption function or number of passes, as these depend on the unique characteristics of the piece and the user's preferences. ImprovNet can be controlled by the user by trying out various combinations of corruption functions at each pass with differing corruption rates. At the global level, users can specify the target genre, context window size, and SSM based preservation ratio, indicating the original segments to be preserved during the refinement process. The specified number of passes pushes each segment towards a target genre as the context surrounding it moves toward it as well.

ImprovNet is able to add chromatic scales, jazz harmonizations, and syncopations in cross-genre classical-to-jazz tasks. However, it has some limitations that we aim to address in future work. The harmonization task sometimes produces overly dense chords, which could be mitigated with additional conditional tokens. Similarly, the onset-duration mask occasionally produces irregular rhythms; we plan to refine this for a more consistent and stable swing rhythm in the future.


\subsection{Generalizability and Link to Diffusion Models}
The corruption-refinement framework used in ImprovNet highlights its potential applicability to broader domains, including large language models (LLM) or other unstructured data tasks, as a generalizable approach for iterative improvement. A single pass of our method shares conceptual parallels with inpainting in Stable Diffusion\footnote{https://stable-diffusion-art.com/inpainting\_basics/}, where corruptions can be interpreted as analogous to the mask applied during the diffusion process. However, the multiple refinement passes in ImprovNet resembles the process of repeating an entire masked diffusion model multiple times (equal to the number of refinement passes), akin to performing several full denoising cycles. However, ImprovNet provides an additional advantage: users can specify the corruption function, allowing for greater controllability over the ``noise" introduced, a level of customization not typically explored in diffusion models. A key distinction lies in the generation process: ImprovNet employs an autoregressive approach, where notes are generated in an ordered sequence, and each segment is iteratively refined, leveraging its temporal context. In contrast, diffusion models generally predict an entire sequence or representation at once, guided by the iterative noise removal process.

\section{Conclusion}
This paper presents ImprovNet, a transformer-based architecture for generating expressive and controllable style-aware musical improvisations, leveraging a self-supervised corruption-refinement training strategy and an iterative generation framework. The model demonstrates strong capabilities in generating both cross-genre and intra-genre improvisations while maintaining structural coherence with the original compositions. ImprovNet's versatility extends beyond style-aware improvisations to include short prompt continuation, short infilling, and genre-conditioned harmonization tasks. Comprehensive objective and subjective evaluations demonstrate ImprovNet's superior performance for the short prompt continuation and short infilling tasks over the Anticipatory Music Transformer baseline. We set the baseline for future work on expressive style-aware improvisations and harmonization.



\bibliographystyle{IEEEtran}
\bibliography{references}

\end{document}